\documentclass[aip,cha,reprint,nofootinbib,superscriptaddress]{revtex4-1}  

\usepackage{amsfonts} 
\usepackage{amssymb}
\usepackage{graphicx}
\usepackage{natbib}
\usepackage{upgreek}
\usepackage{subfigure}

\usepackage[a4paper,left=1.2cm,top=2cm,bottom=2cm,right=1.2cm,foot=1.5cm]{geometry}

\begin{document}
\title{\huge{Manipulating ultracold atoms with a reconfigurable nanomagnetic system of domain walls}}\vspace{-40pt}
\author{A.~D.~West$^1$\noaffiliation}
\author{K.~J.~Weatherill$^1$\noaffiliation}
\author{T.~J.~Hayward$^2$\noaffiliation}
\author{P.~W.~Fry$^3$\noaffiliation}
\author{T.~Schrefl$^4$\noaffiliation}
\author{M.~R.~J.~Gibbs$^2$\noaffiliation}
\author{C.~S.~Adams$^1$\noaffiliation}
\author{D.~A.~Allwood$^2$\noaffiliation}
\author{I.~G.~Hughes$^1$\noaffiliation}
\maketitle

\let\thefootnote\relax\footnotetext{$^1$Physics Department, Durham University, Science Site, South Road, Durham, DH1 3LE, UK. $^2$Department of Materials Science and Engineering, University of Sheffield, Mappin Street, Sheffield, S1 3JD, UK. $^3$Nanoscience and Technology Centre, University of Sheffield, Sheffield, UK. $^4$St. P\"{o}lten University of Applied Sciences, St. P\"{o}lten A-3100, Austria.}

\footnotesize{\textbf{The divide between the realms of atomic-scale quantum particles and lithographically-defined nanostructures is rapidly being bridged. Hybrid quantum systems comprising ultracold gas-phase atoms and substrate-bound devices already offer exciting prospects for quantum sensors\cite{wildermuth,hau}, quantum information\cite{aoki} and quantum control\cite{apjunct}. Ideally, such devices should be scalable, versatile and support quantum interactions with long coherence times. Fulfilling these criteria is extremely challenging as it demands a stable and tractable interface between two disparate regimes. Here we demonstrate an architecture for atomic control based on domain walls (DWs) in planar magnetic nanowires that provides a tunable atomic interaction, manifested experimentally as the reflection of ultracold atoms from a nanowire array. We exploit the magnetic reconfigurability of the nanowires to quickly and remotely tune the interaction with high reliability. This proof-of-principle study shows the practicability of more elaborate atom chips based on magnetic nanowires being used to perform atom optics on the nanometre scale.}

The position, internal state and interactions of quantum particles can be precisely controlled by a variety of techniques utilising combinations of electric, magnetic and optical fields\cite{meystre,adams}. These methods have been enhanced by exploiting the advances in modern nanofabrication techniques, giving rise to a wide array of miniaturised atom chip experiments\cite{atomchips}. Previous studies using micron-scale atom chips have demonstrated robust and exquisite control over atoms through increasingly complex networks of traps, guides and other atom-optical elements. Further miniaturisation of such devices to the nanoscale offers the tantalising prospect of the precise manipulation of the position and internal state of individual atoms. 

Magnetic atom chips can be roughly divided into devices based on current-carrying wires\cite{zimm} and those based on permanent magnetic material\cite{hindshughes}. Atom chips based on current-carrying wires can suffer from technical noise which induces spin-flip losses\cite{mattnoise} and causes inhomogeneities in the magnetic potentials. Care must also be taken to ensure sufficient power dissipation, which can limit the precision of the fields created. On the other hand, lithographically fabricated permanent magnets far surpass the feature size limits of atom chips based on current-carrying wires\cite{westervelt,lev,gdtbfeco} and offer greater flexibility of design, whilst allowing the creation of significantly stronger fields. This has enabled the creation of atom traps with exceptionally high trap frequencies\cite{fept,spreeuw,hannaford}. However, permanent magnets suffer from the inability to be switched off or reconfigured once the device has been fabricated, limiting the realisation of dynamic behaviour. 
\begin{figure}[!t]
\subfigure{\includegraphics[scale=0.43]{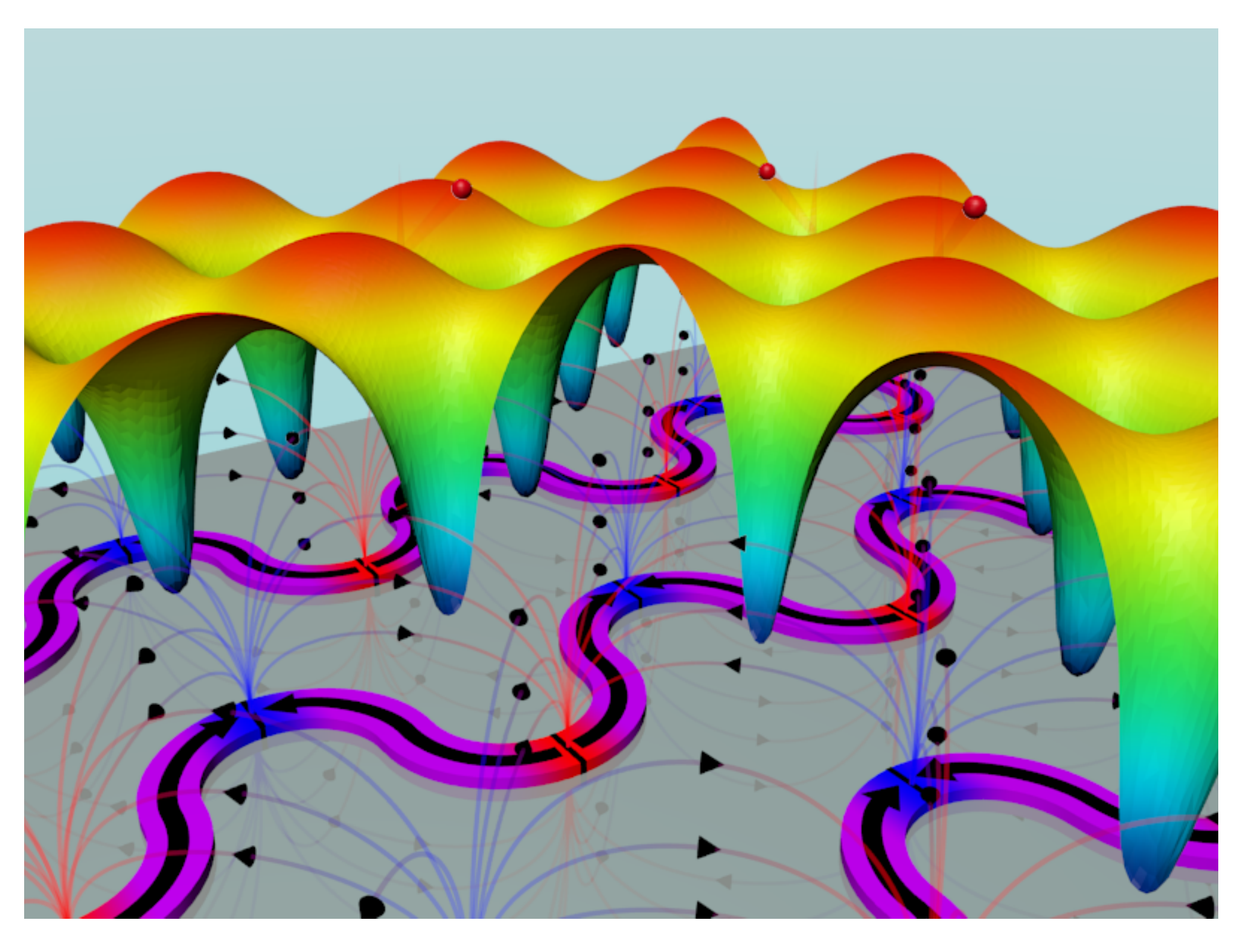}}\vspace{-8pt}
\subfigure{\includegraphics[scale=0.32]{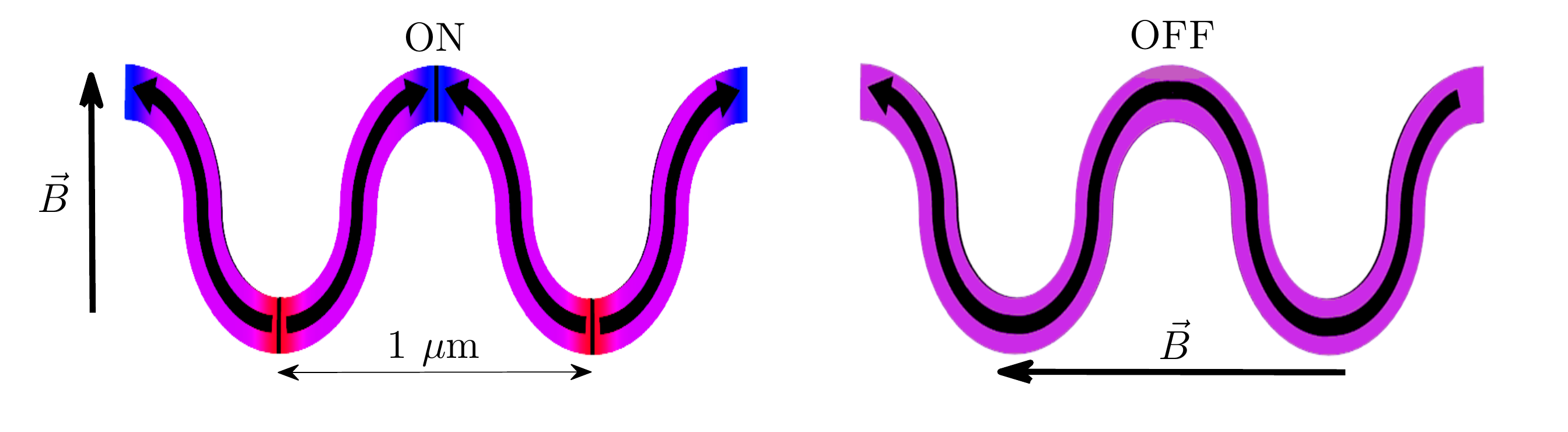}}
\caption{\scriptsize{A schematic representation of the experiment. The array of undulating nanowires is shown. The wires are 125~nm wide and 30~nm thick and have a periodicity of 1~$\upmu$m. The wire shading represents the magnetic polarity; the wires are shown in the `on' state, hosting DWs at each apex. Also indicated are magnetic field lines emanating from and entering the DWs. Above the wires is shown the magnetic potential isosurface corresponding to a value of $|\vec{B}|=1.57$~mT -- shading indicates the height. The two magnetisation configurations of the nanowires are illustrated beneath: the `on' state with many domains and the `off' state with one domain and no DWs. Also labelled are the directions of magnetic field required to switch into these states (directions antiparallel to these are equivalent).}}
\label{fig:schematic}
\end{figure}

Here we demonstrate an atom chip based on nanomagnetic technology that exhibits the benefits of patterned magnetic materials whilst maintaining reconfigurability. The small characteristic size of our magnetic nanostructures provides exquisite control over the magnetic configuration and, therefore, the atomic interaction with our device. We have interfaced ultracold atoms with the fringing fields from an array of 180$^{\circ}$ DWs in magnetic nanowires. The head-to-head type DWs found in planar magnetic nanowires have an associated magnetic monopole moment and produces fringing fields which are ideal for manipulating paramagnetic atoms\cite{nanoapl}. The creation and position of DWs can be accurately and reliably controlled by the choice of nanowire geometry and subsequently manipulated via the application of external magnetic fields, currents or stress\cite{danstress}. Nascent spintronic technologies\cite{wolf}, such as DW logic devices\cite{domainlogic} and `racetrack' memory\cite{racetrack}, have already demonstrated the utility of DWs in creating intricate data networks. We use the same technology to generate and modify the detailed magnetic field pattern from our chip, thus providing a controllable atom-field interaction. 

Planar permalloy (Ni$_{80}$Fe$_{20}$) nanowires exhibit a magnetic easy axis along their length. The wires we produce have an undulating shape which results in the magnetisation configuration having two distinct states, as shown in Figure~\ref{fig:schematic}. The ground or `off' state contains a single continuous magnetic domain along the length of the wire. In this state there are nominally no out-of-plane fringing fields and thus no interaction with the atoms. A higher energy, metastable `on' state is also possible, hosting many domains, situated at the wire apexes. This results in large fringing fields and hence a strong interaction. The nanowires can be forced into the `on' state by application of an external in-plane magnetic field pulse, orthogonal to the length of the wire to saturate magnetisation in this direction. Relaxation after the pulse results in the magnetisation aligning to local wire edges, leading to a DW forming at each wire apex. The result is an extended periodic 2D array of eight million alternating north and south poles\cite{nanojap} which constitutes a magnetic mirror which reflects atoms\cite{hindshughes}. The ground state can then be recovered via the application of a magnetic field pulse along the length of the wire, causing pairwise annihilation of the DWs and leaving a continuous magnetisation.
\begin{figure*}[!t]\hspace{-10pt}
\includegraphics[scale=0.56]{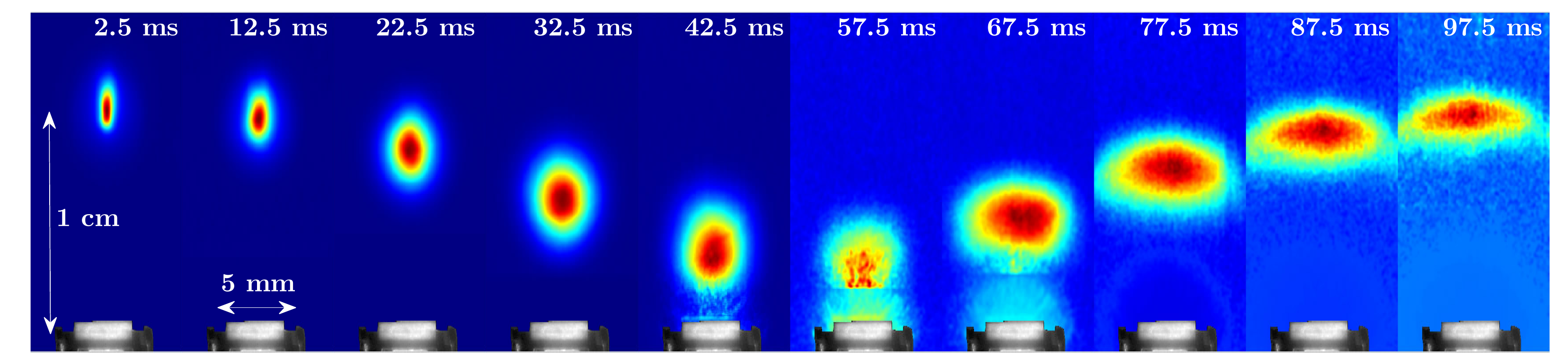}
\caption{\scriptsize{A series of fluorescence images of the atomic cloud as they are reflected from the nanowire array. Superimposed is an image of the chip and surrounding mount. For later times the the colour map is manually and locally rescaled to suppress the visibility of significant scattered light from the chip mount.}}
\label{fig:bounceseq}
\end{figure*}

A paramagnetic atom entering a magnetic field experiences a Zeeman interaction energy, $E_{\rm Z}$, given by
\begin{equation}
E_{\rm Z}=-\vec{\mu}.\vec{B}=m_Fg_F\mu_{\rm B}|\vec{B}|,
\end{equation}
where $m_F$ is the atom's magnetic quantum number, $g_F$ is the Land\'{e} g-factor and $\mu_{\rm B}$ is the Bohr magneton. Thus a magnetic field gradient will result in the atom experiencing a Stern-Gerlach force, $F_{\rm SG}$, given by $F_{\rm SG}=-\vec{\nabla}E_{\rm Z}$. Atoms which have $m_Fg_F>0$ are `weak-field-seeking' and are repelled from high magnetic fields. In our experiment a cloud of $^{87}$Rb atoms is laser-cooled at a height of 10~mm above the nanowire array in a magneto-optical trap, and is optically pumped to ensure that it is prepared in a weak-field-seeking ($\left|F=2, m_F=2\right>$) state. The cloud is then allowed to fall under gravity such that the atoms enter the fringing fields from the DWs and are subsequently reflected, observed in a series of fluorescence images shown in Figure~\ref{fig:bounceseq}. This clearly demonstrates the feasibility of manipulating ultracold atoms using DW fringing fields. 

The atom dynamics during reflection are determined by the shape and magnitude of the fringing fields, which have been calculated. Such calculations show that extremely large magnetic field gradients are produced; within 10~nm of the surface these can reach $\sim 10^6$~T/m with corresponding field strengths of around 1~T. An atom dropped from a height of 10~mm will be reflected at heights of \textless 500~nm. The field gradients in this region are sufficiently large that an atom approaching the nanowires effectively experiences a point interaction with the field. Thus it is possible to use the calculated fields to define an effective isosurface of $|\vec{B}|$ from which the atoms can be considered to reflect specularly. This surface is a level set of the field (1.54~mT) for which the Zeeman interaction energy of a $^{87}$Rb atom in the specified magnetic sublevel is equal to the initial gravitational potential energy. This surface is illustrated above the nanowire array in Figure~\ref{fig:schematic}. An ideal magnetic mirror has a sinusoidally varying in-plane magnetisation\cite{hindshughes}, which yields a flat equipotential surface. The 2D array of DWs mimics the general pattern of alternating magnetisation direction, but deviations from a pure sinusoidal form produce a roughness in $|\vec{B}|$, and hence a corrugated isosurface. The result is that a cloud of atoms entering the DW fringing fields will be reflected diffusely. This is shown in Figure~\ref{fig:bounceseq} by the increased lateral extent of the cloud after reflection.

Quantitative analysis of the dynamics of the atoms is provided by the use of an intensity-stabilised light sheet, as illustrated in Figure~\ref{fig:setup}. A resonant laser beam is focussed in one direction by cylindrical lenses, passing between the nanowire array and the initial location of the atomic cloud. As falling atoms pass through the light sheet they scatter photons, causing a dip in intensity. After the atoms have been reflected there is a second dip as they pass through again on the way back up. The light sheet has low intensity and is retroreflected to minimise any perturbation of the atoms. We estimate a sensitivity of around $5\times 10^3$ atoms with negligible heating of the cloud.
\begin{figure}[!h]
\subfigure{\includegraphics[scale=0.39]{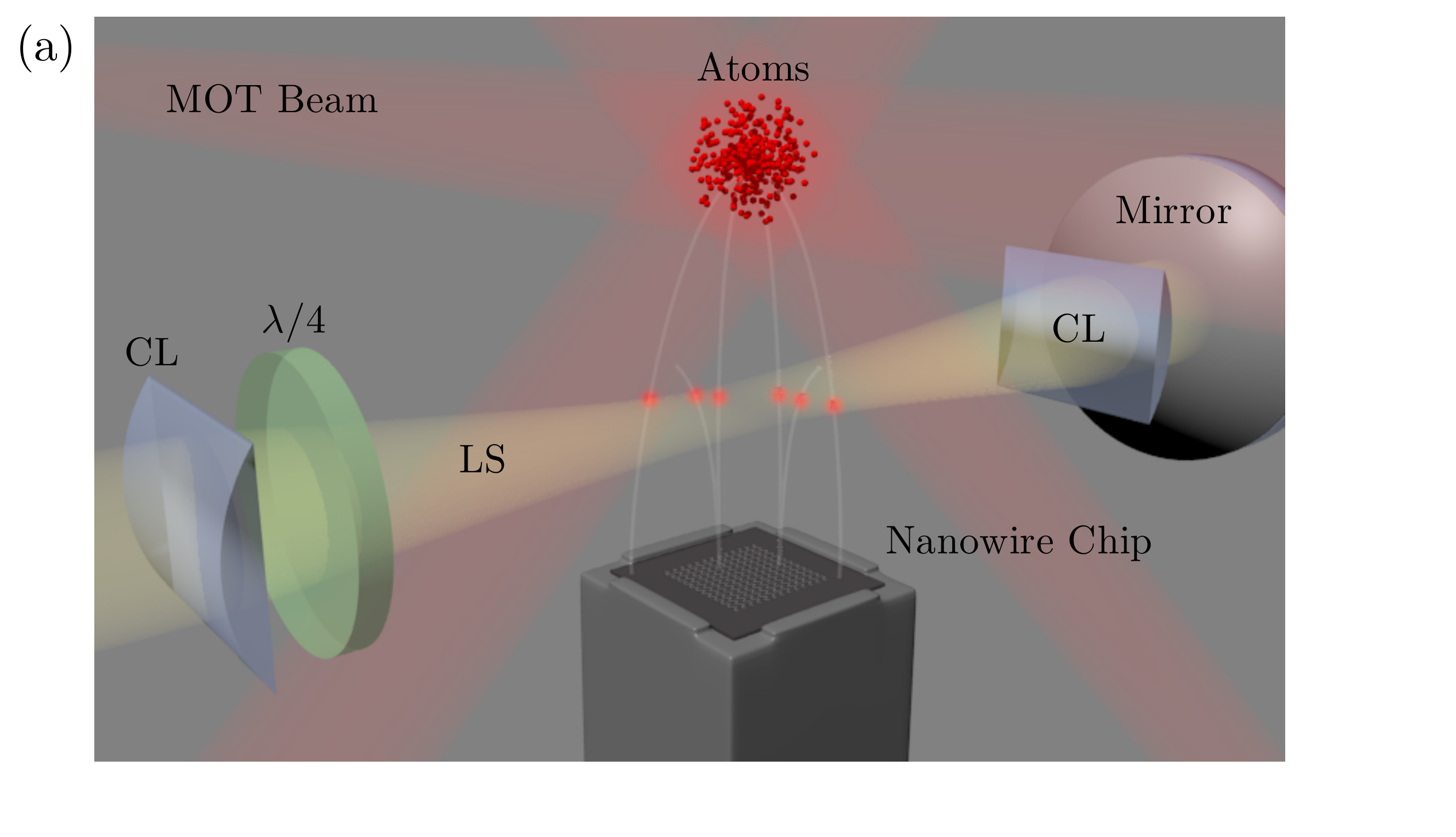}}\vspace{-15pt}
\subfigure{\includegraphics[scale=0.35]{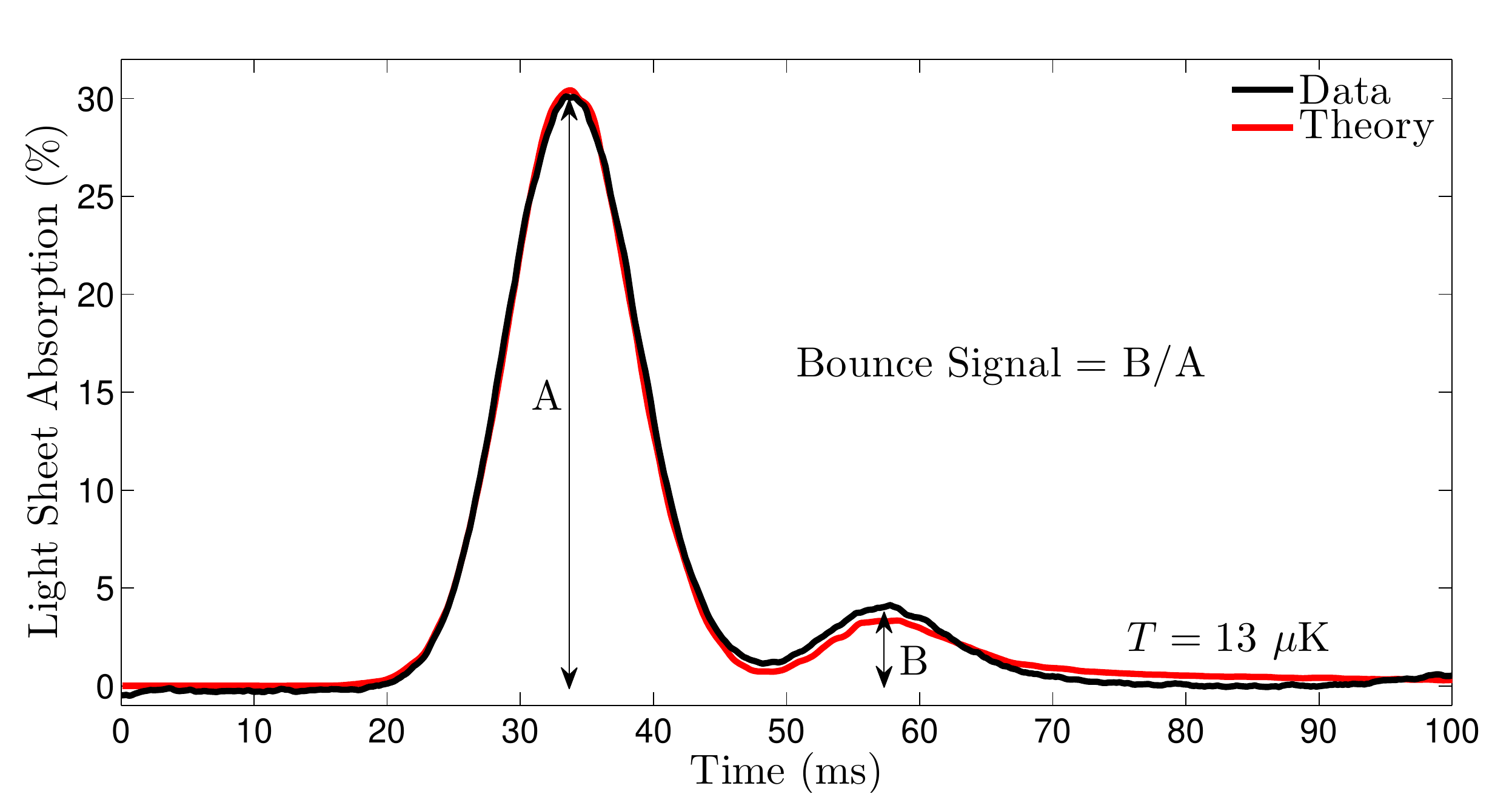}}
\vspace{-12pt}
\caption{\scriptsize{(a) An illustration of the experimental setup -- a cloud of atoms is laser-cooled and optically pumped at a height of 10~mm above the nanowire array and then allowed to fall under gravity. A resonant laser beam is focussed in one direction by a cylindrical lens (CL) and polarised by a quarter waveplate ($\lambda$/4) before being retroreflected and then detected by a photodiode. The atoms pass through this light sheet onto a 2~mm~$\times$~2~mm written array, from which they are reflected and subsequently pass through the light sheet for a second time. (b) An example signal obtained from the light sheet for a cloud initially at 13~$\upmu$K. The first, taller peak corresponds to atoms falling through the light sheet. The second peak corresponds to atoms reflected from the nanowire array. We define the bounce signal as the ratio between the height of these two peaks. Also plotted is the prediction from Monte Carlo simulations.}}
\label{fig:setup}
\end{figure}
An example of the signal provided by the light sheet is shown in Figure~\ref{fig:setup}(b). Also shown is the theoretical prediction made via Monte Carlo simulations of the atom cloud dynamics. We see excellent agreement between theory and data for a large range of parameters. The bounce signal is strongly dependent on initial temperature; if the cloud is too hot then the atoms spread out such that there are fewer atoms in the light sheet at any time, reducing the overall signal size, and the two peaks broaden such that it is no longer possible to resolve the bounce signal. For this reason, the remainder of the work was conducted at our minimum temperature of 10-15~$\upmu$K. 

The `on' and `off' magnetic states of the nanowires (c.f. Figure~\ref{fig:schematic}(b)) are easily selected using single magnetic field pulses. Thus the fringing fields can be toggled, and the mirror switched on and off as the atom-field interaction is tuned. Investigation of this switching process was carried out via examination of the bounce signal, which represents a probe of the collective magnetic configuration. The resulting data are shown in Figure~\ref{fig:onoff}(a). We observe that the interaction can be toggled with 100\% reliability. The pulses used to reconfigure the array have a minimum duration of around 0.5~ms, limited by the inductance of the coils, but in theory could be as short as a few ns, as the switching process occurs on an extremely short timescale\cite{dwdynamics}. Varying the size of the field used to switch the mirror permits analysis of the micromagnetic behaviour of the array, as the reflectivity of the device is directly correlated to the resulting bounce signal. Example data resulting from this process are shown in Figure~\ref{fig:onoff}(b). 
\begin{figure}[!h]
\includegraphics[scale=0.32]{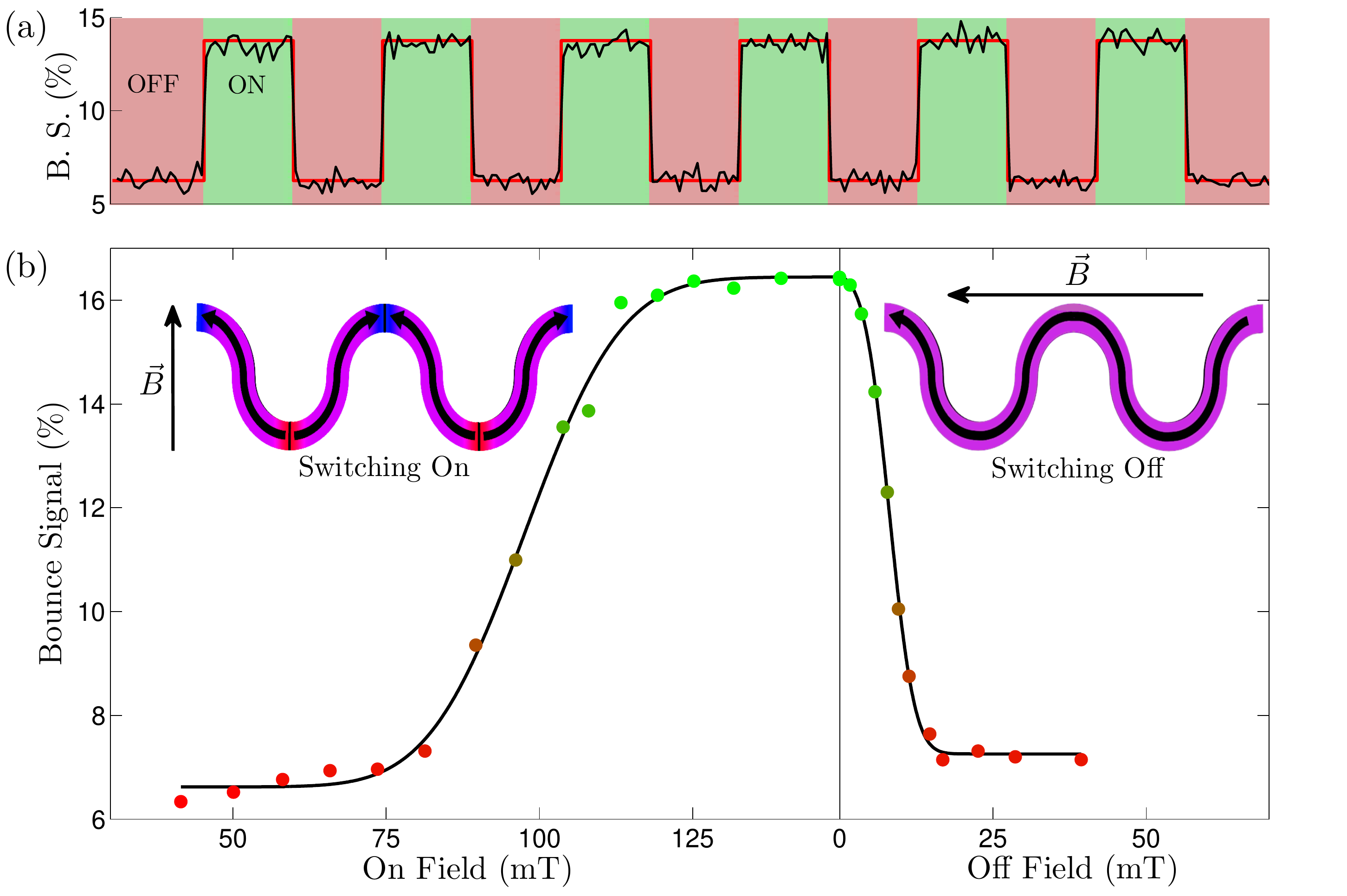}
\caption{\scriptsize{(a) Repeated switching of the array. Single shot measurements of the dynamics are recorded 20 times followed by single application of a switching field pulse. The bounce signal as defined in Figure~\ref{fig:setup}(b) is extracted from each measurement. The red line is a guide to the eye. (b) Dependence of the bounce signal on the size of the field used to switch the magnetisation state of the nanowire array. The insets illustrate the two magnetisation states of the nanowires as per Figure~\ref{fig:schematic}. The error bars are smaller than the size of the data points. The black line is a fit of the form of an error function.}}
\label{fig:onoff}
\end{figure}
There is a gradual turning `on' and `off' of the chip, because the micromagnetic reconfiguration processes of populating and annihilating DWs occur over ranges of applied external field. The data are fit very well (reduced $\chi^2\approx 1$) by error functions, which indicates that there is a normal distribution of required switching fields amongst the DW sites. This inherent variation indicates random fluctuations within the fabrication process and the stochastic nature of magnetic switching events. From the fits we extract a mean switching-on field, of 98.0 $\pm$ 0.4~mT with a width, $\sigma$, of 15.8 $\pm$ 0.8~mT. The mean switching-off field is 8.1 $\pm$ 0.3~mT with a width of 4.6 $\pm$ 0.4~mT. Measurements using the magneto-optical Kerr effect and scanning Hall probe microscopy are in good agreement with the fits in Figure~\ref{fig:onoff}. In normal operation fields significantly larger than the mean values are used to try and ensure switching of the entire array. 

We have experimentally demonstrated the manipulation of ultracold atoms using DWs in magnetic nanowires, thus establishing the feasibility of creating atom chip devices based on reconfigurable nanomagnetic technology. Via a detailed examination of the interface between these two disparate regimes we have also shown that it is possible to use atom dynamics as a probe of nanomagnetic behaviour, which could lead to immediate applications. For example, experiments similar to the one we present could be used to probe artificial spin-ice arrays\cite{spinice} with different magnetic monopole configurations (and therefore fringing fields) but near-identical overall magnetisation. Work is also currently underway to exploit the inherently small lengthscale of the device to allow surface interactions to be studied at distances of less than 50~nm, c.f.\cite{ashok}.

However, the most attractive feature of the approach exemplified by our proof-of-principle experiment is that it allows for the control of atomic motion via magnetic interactions that can be quickly, remotely and reliably tuned through the application of macroscopic magnetic fields. The intrinsic freedom afforded by the lithographic methods used to create planar nanowires, as well as the inherently mobile nature of the DWs, opens the door to a variety of devices. For example, DW fringing fields have already been utilised to confine macroscopic objects\cite{beads,danbeads}, and we plan to use them in an analogous manner as the basis for mobile atom traps\cite{nanoapl}. A scheme for atom trapping based on such nanomagnetic technology would provide exceptional atomic confinement through extremely high trap frequencies and trapping volumes smaller than the de Broglie wavelength\cite{folman}. When combined with the controlled transport of DWs this will offer a dynamic architecture that lends itself well to future quantum information processing applications.

\section*{Methods Summary}
\footnotesize{
Details of the design, manufacture and characterisation of the nanowire array can be found in \cite{nanojap}. Theoretical calculations of the DW fringing fields were performed using a simple analytical model described in \cite{mono}. The accuracy of the model was verified through comparison with proprietary code which solves the Landau-Lifshitz-Gilbert equation within a finite element/finite boundary framework \cite{schrefl}. We observe errors less than 10\% for distances above 100~nm from the nanowire surface. The effective isosurface is calculated from these fringing fields.

Predictions of the atom dynamics were provided by Monte Carlo simulations which propagate classical trajectories of non-interacting atoms. The distributions of atom position and velocity are initialised in accordance with experimental data. Repulsion from the fringing fields is treated as a classical reflection from the computed effective isosurface. The effects of the optical pumping and light sheet beams on the internal dynamics and motion of the atoms are included. The only free parameter is atom number.

We prepare the state of the nanowire array using a set of Helmholtz coils. Up to 200~A is passed through the coils by discharging car batteries. A rheostat allows a continuous range of magnetic fields to be selected. The `on' coils are also used to realise the field for the magneto-optical trap (MOT). The MOT is loaded for 10~s followed by optical molasses, cooling the atoms to around $10~\upmu$K. The atoms are then optically pumped to the desired state. A quantisation axis is maintained by a magnetic field of 0.2~mT for the duration of the experiment. The atoms pass through a circularly polarised, retroreflected light sheet which is intensity-stabilised to better than 0.1\%. The $1/e^2$ radii of the light sheet beam waist are 90~$\upmu$m and 3.9~mm. The signal from the light sheet is monitored using an amplified avalanche photodiode.}

\bigskip

\scriptsize{
\noindent\textbf{Acknowledgements}This work was supported financially by EPSRC grants EP/F025459/1 and EP/F024886/1.
\bigskip

\noindent\textbf{Author Contributions}A.D.W., K.J.W., T.J.H., D.A.A. and I.G.H. contributed to the writing of the manuscript, the experiment was designed and built by A.D.W. and K.J.W., data analysis was carried out by A.D.W., K.J.W. and I.G.H., Monte Carlo simulations were performed by A.D.W., micromagnetic simulations were performed by T.J.H. and T.S., the magnetic nanostructures were designed by T.J.H., D.A.A., M.R.J.G. and T.S., fabrication and characterisation of the nanowires was carried out by T.J.H. and P.W.F., C.S.A. came up with the initial concept, D.A.A. and I.G.H. managed the project.
\bigskip

\noindent\textbf{Author Information} The authors declare no competing financial interests.
Correspondence and requests for materials should be addressed to I.G.H. (i.g.hughes@durham.ac.uk).}

\vspace{20pt}

\section*{Methods}
\scriptsize{
The nanowires were fabricated using electron-beam lithography with liftoff processing. Metallisation was achieved by thermal evaporation. A 2~mm~$\times$~2~mm area was written with $2\times 10^3$ ${\rm Ni}_{80}{\rm Fe}_{20}$ nanowires of cross section 125~nm~$\times$~30~nm. The wires are written with an undulating pattern of wavelength 1~$\upmu$m and adjacent wires are displaced by a distance of 1~$\upmu$m. All the wires extend across the full 2~mm of the written area.

Characterisation of the magnetisation reconfiguration processes of the wires was performed previously via magnetic hysteresis measurements using MOKE\cite{nanojap}. This gives a measurement of the collective reconfiguration of the magnetisation structure. Additional characterisation was also provided by SHP microscopy\cite{nanojap} which provides imaging of individual DWs over a very small fraction of the chip.

Theoretical calculations of the resulting fringing fields created by the DWs were performed using a simple analytic model which we have developed\cite{mono}. In its most basic form, the DW is approximated as a point `magnetic charge', $q_{\rm m}$, which has a size proportional to the wire's cross-section. The magnetic flux density associated with this charge is then given by $\vec{B}(\vec{r})=q_{\rm m}/4\pi r^2\ \hat{r}$. Extending the model allows the DW to be represented as a 2D object, affording a more accurate calculation. The accuracy of this model was verified through comparison with calculations based on micromagnetically simulated magnetisation structures. This method utilises proprietary code to solve the Landau-Lifshitz-Gilbert equation within a finite element/finite boudnary framework\cite{schrefl}. The resulting fringing fields can be computed from the equivalent dipole charges at the surface of the nanowire. We observe excellent accuracy when using the analytic models with errors typically less than 10\% for heights above 100~nm.

Tesselating the result from our analytic model yields the field from the domain wall array and hence allows for computation of the effective isosurface. The roughness of the isosurface can be characterised by the mean zenith angle, i.e. the angle between 0$^{\circ}$ and 90$^{\circ}$ quantifiying the deviation from a flat surface. This was calculated to be $28.8^{\circ}$. Also of note is the fact that there are small regions between DWs where there is insufficient field to reflect the atoms; atoms impinging on these regions will either scatter stochastically from, or be adsorbed onto, the substrate.

Quantitative predictions of the atom dynamics are provided by Monte Carlo simulations which propagate classical trajectories of non-interacting atoms. The initial distributions of atom position and velocity are populated in accordance with experimental data. Repulsion from the fringing fields is treated as a classical reflection from the computed effective isosurface. The effects of the optical pumping and light sheet beams on the internal dynamics and motion of the atoms, and hence the resulting light sheet signal, are also included. The only free parameter in the simulations is atom number. Fitting this parameter is in fact a good method of measuring atom number as it is proportional to the integrated signal.

In our experimental procedure we initially prepare the configuration of the nanowire array by pulsing currents of up to around 200~A through Helmholtz coils in vacuum. The required current is provided by discharging a set of car batteries in series. Use of a rheostat allows a continuous range of fields to be selected. The `on' coils are also used to realise the magneto-optical trap (MOT) which is loaded for 10~s and followed by a molasses ramp, cooling the atoms to around $10~\upmu$K. Modifying this molasses ramp reliably changes the resultant cloud temperature whilst maintaining a consistent experimental sequence. The atoms are then optically pumped for a duration of 2~ms, during which a quantisation axis parallel to the optical pumping and light sheet beams is defined via application of an external field of around 0.2~mT. This quantisation axis is kept on as the atoms fall in order to minimise any depolarisation of the atoms caused by the light sheet, and to ensure the adiabaticity of the atoms' motion as they enter the large fringing fields. Optical pumping is achieved by addressing the $F=2\rightarrow F^{\prime}=2$ transition with a laser beam with 0.6~$\upmu$W of power, and the $F=1\rightarrow F^{\prime}=2$ transition with 40~$\upmu$W, both beams have a $1/e^2$ radius of 1.4~mm. The atoms pass through the retroreflected light sheet which is intensity-stabilised with a power of 50~nW. The $1/e^2$ radii of the semi-major axes of the light sheet's profile are 90~$\upmu$m and 3.9~mm. Both the optical pumping beams and the light sheet beam are circularly polarised by a single quarter-wave plate. The light level is monitored using a Hamamatsu C5460 APD module.}

\end{document}